# Bound States in the Compactified Gravity

Askold Duviryak

Institute for Condensed Matter Physics of the National Academy of Sciences of Ukraine, Lviv

**ABSTRACT**

A potential of pointlike mass in the partially compactified multidimensional space is considered. The problem is reduced to the multidimensional Poisson equation with the Dirac comb source in r.h.s. Explicit solutions are built in the cases of 2D and 4D spaces with one compact dimension. The last example of the potential is used in the Schrödinger equation. Bound states of a gravitating test particle on 3D brane of 4D compactified space are studied by means of various approximate methods.

## 1 INTRODUCTION

Attempts to unify the gravitation with other fundamental interactions have led to the idea that the space is more than three-dimensional (and the space-time is more than four-dimensional) (Duff, *etel* 1986). Shortly after the discovery of the general relativity, Kaluza and Klein unified the electrodynamics and gravitation in terms of 5D space-time. Einstein and other theorists assumed that a redundant dimension is compact, of very short extension, e.g., of order $R \approx \ell_{Pl} = \sqrt{G\hbar/c^3} \approx 1.6 \cdot 10^{-33}$ cm, so that a resolution of physical devices is not sufficient to detect this dimension in experiments.

As Kaluza-Klein and other unified theories developed (Appelquist, *etel* 1987), the number of hidden dimensions grew. A similar picture is characteristic of the modern superstring theory (Marshakov, 2002) where the space is $N = 9$ dimensional (hereinafter $N$ does not count a time dimension). All the conventional matter is attached to $D < N$ dimensional submanifold in the space, but the gravity (and maybe other unknown kinds of "sterile" matter) can penetrate into redundant $N - D \equiv d$ dimensions (9 – 3 = 6 in our case) which are very compact and thus invisible.

Slightly different picture is given rise within the ADD-hypothesis based on the string theory (Arkani-Hamed, *etel* 1998; Rubakov, 2001). There the extension of the compact dimensions must not be very short. Since all the matter (but the gravity) is 3D, the extra dimensions may show up only via certain gravitation effects on microscopic or mesoscopic levels. A black hole creation in LHC (Krasnikov & Matveev, 2004) could be the first expected example.

Explanation of how a multidimensionality influences the gravitation does not need necessarily a framework of the string theory, Kaluza-Klein theory or even general relativity. In the present paper, the effect of extra compact dimensions on the gravity is demonstrated within the Newton theory. For this purpose, solutions of the Poisson equation with pointlike source in partially compactified spaces are considered. Low-dimensional cases are studied. Explicit solutions are built in the cases of 2D and 4D spaces with one compact dimension. The latter potential is used in the quantum Kepler-Coulomb problem. The corresponding Schrödinger equation is solved by means of various analytical approximations and the numeral integration. Physical consequences of the derived solutions are discussed.

## 2 POISSON EQUATION IN THE COMPACTIFIED SPACE

Let us consider $N$-dimensional Euclidean space $\Re^N$ which is infinite in all dimensions. Points of this space are parameterized by means of Cartesian coordinates $\mathbf{z} = \{z_1,\ldots, z_N\}$. A pointlike mass $M$ generates in this space the gravitational potential satisfying the $N$-dimensional Poisson equation:

$$\Delta^{(N)}\varphi(\mathbf{z}) = \Omega_{(N)} G_{(N)} M \delta^{(N)}(\mathbf{z}), \qquad \text{where} \quad \Delta^{(N)} = \frac{\partial^2}{\partial z_1^2} + \cdots + \frac{\partial^2}{\partial z_N^2} \qquad (2.1)$$

and $\delta^{(N)}(\mathbf{z})$ are the $N$-dimensional Laplacian and the Dirac $\delta$-function,

$$\Omega_{(N)} = 2\pi^{N/2} \big/ \Gamma(N/2) \qquad (2.2)$$

is the area of a unit hypersphere, and $G_{(N)}$ is the gravitational constant in $\Re^N$. The solution of Eq. (2.1) is known (Ivanenko & Sokolov, 1953):

$$\varphi(\mathbf{z}) = -\frac{G_{(N)} M}{(N-2) z^{N-2}}, \qquad (2.3)$$

where $z = |\mathbf{z}|$. It is valid for $N \geq 3$ and reduces to Newton (or Coulomb) potential for $N = 3$.

### 2.1 General case of compactification: $N = D + d$

Let us now the space is infinite in part of dimensions, and is compact in other ones. In the simplest case such space is $(N = D + d)$-dimensional manifold $\boldsymbol{M}_{D+d} = \Re^D \times \mathbf{T}_d \equiv \Re^D \times \mathbf{S}_1 \times \cdots \times \mathbf{S}_1$, where factors $\mathbf{S}_1$ of $d$-dimensional torus $\mathbf{T}_d$ are assumed for simplicity to be circles of the same *compactification radius* $R$. Hereinafter the parameterization is used: $\mathbf{x} = \{x_1,\ldots, x_D\}$, $x_i \in (-\infty, \infty)$ of $\Re^D$, and $\mathbf{y} = \{y_1,\ldots, y_d\}$, $y_j \in [0, 2\pi R)$ of $\mathbf{T}_d$.

The differential equation (2.1) is a local form of the gravity (or Coulomb) law, and no global features of the space are involved in (2.1). Thus, this equation is applicable in the case of a compactified space. A topology of the manifold $\mathcal{M}_{D+d}$ is taken into account only via global properties of functions involved in the equation (2.1). In particular, $\delta^{(N)}(\mathbf{z}) = \delta^{(D)}(\mathbf{x})\widetilde{\delta}^{(d)}(\mathbf{y})$, where $\delta^{(D)}(\mathbf{z})$ is a usual $D$-dimensional $\delta$-function, but $\widetilde{\delta}^{(d)}(\mathbf{y})$ is a $d$-dimensional Dirac comb (Flügge 1971), $2\pi R$-periodical in each dimension:

$$\delta^{(D)}(\mathbf{x}) = \frac{1}{(2\pi)^D}\int d^D k\, e^{i\mathbf{k}\cdot\mathbf{x}}, \qquad \widetilde{\delta}^{(d)}(\mathbf{y}) = \frac{1}{(2\pi R)^d}\sum_{\mathbf{q}} e^{i\mathbf{q}\cdot\mathbf{y}};$$

here $\mathbf{q} = \{n_1/R,\ldots, n_d/R\}$, $n_i = 0, \pm 1, \pm 2,\ldots$ Similarly, the solution $\varphi(\mathbf{x},\mathbf{y})$ of the equation (2.1) can be presented via the Fourier integral and series:

$$\varphi(\mathbf{x},\mathbf{y}) = -\frac{\Omega_{(D+d)}}{(2\pi)^{D+d} R^d}\sum_{\mathbf{q}}\int d^D k\, e^{i(\mathbf{k}\cdot\mathbf{x}+\mathbf{q}\cdot\mathbf{y})}\frac{G_{(D+d)}M}{\mathbf{k}^2 + \mathbf{q}^2}. \qquad (2.4)$$

When considering the asymptotics at $|\mathbf{y}| \ll R$, the summation step $|\Delta\mathbf{q}| \sim 1/R$ can be regarded as small one. Thus, a summation is closed to an integration:

$$\frac{1}{R^d}\sum_{\mathbf{q}} f(\mathbf{q}) \to \int d^d q\, f(\mathbf{q}), \text{ and yields } \varphi(\mathbf{x},\mathbf{y}) \to -\frac{G_{(D+d)}M}{(D+d-2)(\mathbf{x}^2+\mathbf{y}^2)^{\frac{D+d}{2}-1}},$$

the potential (2.3) for the infinite ($N=D+d$)-dimensional space. So physically, a compactness of the space $\mathcal{M}_{D+d}$ is not seen in the scale $\ll R$. In order to calculate the asymptotics at $|\mathbf{x}| \gg R$ we integrate r.-h.s. of Eq. (2.4) over $\mathbf{k}$:

$$\varphi(\mathbf{x},\mathbf{y}) = -\frac{\Omega_{(D+d)} G_{(D+d)}M}{(2\pi)^{D/2}(2\pi R)^d}\sum_{\mathbf{q}} e^{i\mathbf{q}\cdot\mathbf{y}}\left(\frac{q}{x}\right)^{\frac{D}{2}-1} K_{\frac{D}{2}-1}(qx), \qquad (2.5)$$

then take into account the condition $qx \gg 1$, except of the case $q = 0$, and use the properties $K_s(\xi) \underset{\xi\to\infty}{\to} \sqrt{\frac{\pi}{2\xi}}e^{-\xi}$, $\xi^s K_s(\xi) \underset{\xi\to 0}{\to} 2^{s-1}\Gamma(s)$ of Macdonald function (Ivanenko & Sokolov, 1953) in r.-h.s. of Eq. (2.5). All terms of the sum (2.5) are exponentially small except that of $\mathbf{q} = 0$. Thus, we arrive at the result:

$$\varphi(\mathbf{x},\mathbf{y}) \underset{x>>R}{\to} -\frac{\Omega_{(D+d)}}{\Omega_{(D)}(2\pi R)^d}\frac{G_{(D+d)}M}{(D-2)x^{D-2}} = \varphi(\mathbf{x}) \qquad (2.6)$$

which does not depend on the coordinates **y** of compact dimensions. Even more, it coincides formally with the potential (2.3) of a pointlike mass $M$ in an infinite $D$-dimensional space with the effective gravitational constant

$$G_{(D)} = \frac{\Omega_{(D+d)}}{\Omega_{(D)}(2\pi R)^d} G_{(D+d)}. \qquad (2.7)$$

Thus, compact dimensions **T**$_d$ of the space **M**$_{D+d}$ are gravitationally invisible in the scale $>> R$.

## 2.2 The case of compactified plane: $N = 1 + 1$

The compactified plane **M**$_{1+1} = \Re \times$**S**$_1$ is parameterized by variables $x \in (-\infty, \infty)$ and $y \in [0, 2\pi R)$. The potential (2.4) turns into a formal series:

$$\varphi(x,y) = -\frac{G_{(2)}M}{2\pi R}\sum_{n=-\infty}^{\infty}\int_{-\infty}^{\infty}dk\,\frac{e^{i(kx+ny/R)}}{k^2+(n/R)^2} = -\sum_{n=-\infty}^{\infty}\frac{G_{(2)}M}{2|n|}e^{-\frac{|nx|-iny}{R}} \equiv \sum_{n=-\infty}^{\infty}\varphi_n(x,y)$$

The term $\varphi_0(x,y)$ of this sum is divergent, but it can be regularized by extracting an infinite constant $\varphi_0(0,0)$ so that the difference $\varphi_0(x,y) - \varphi_0(0,0) \propto |x|/R$ is finite. The rest of the sum yields complex conjugated terms, thus the regularized potential

$$\varphi(x,y) = \frac{G_{(2)}M}{2}\left\{\frac{|x|}{R} + \ln\left(1-e^{-\frac{|x|+iy}{R}}\right) + \ln\left(1-e^{-\frac{|x|-iy}{R}}\right)\right\} \qquad (2.8)$$

is real; see Figure 1. Close to the source, where $\rho = |x+iy| << R$, the function (2.8) has the asymptotics $\varphi(x,y) \to G_{(2)}M \ln(\rho/R)$, the potential in an infinite plane. The limit $|x| >> R$ leads to the uniform field $\varphi(x,y) \to \tfrac{1}{2}G_{(2)}M|x|/R$.

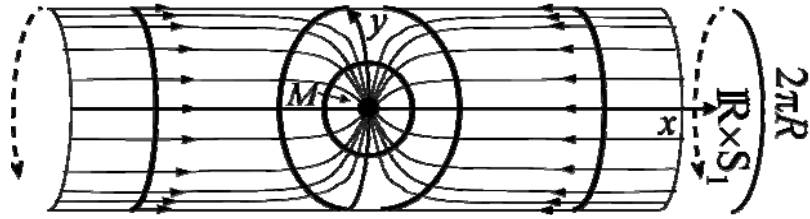

**Figure 1:** Potential $\varphi(x,y)$ of pointlike source $M$ on the cylinder $\mathcal{M}_{1+1} = \Re \times \mathbf{S}_1$:
→ tension lines; — equipotential 1D surfaces.

Noteworthy the potential (2.8) admits an electrotechnical treatment. If one lets variable $y$ run all over the real axis $\Re$, the potential turns periodical in $y$.

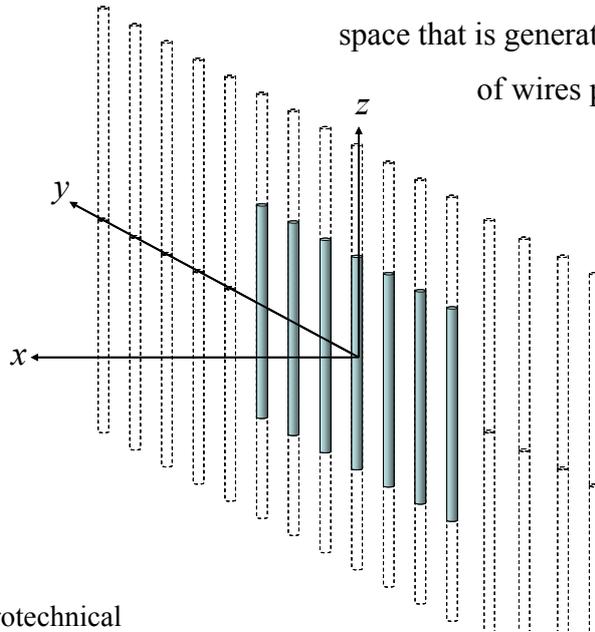

It then describes the electrostatic field in 3D (infinite) space that is generated by the infinite lattice of wires placed in the plane $x = 0$ parallelly to the axis $Oz$ with the step $\Delta y = 2\pi R$ and charged up to

$$\frac{\Delta q}{\Delta z} = -\frac{G_{(2)}M}{2}$$

per unit length; see Figure 2.

**Figure 2:**
The electrotechnical treatment of the potential (2.8).

## 2.3 The case $N = 3 + 1$

The compactified plane $\mathcal{M}_{3+1} = \Re^3 \times \mathbf{S}_1$ is important physically since it is related to the Kaluza-Klein theory. An evaluation of potential (2.4) is straightforward:

$$\varphi(\mathbf{x},y) = -\frac{G_{(4)}M}{4Rr}\sum_{n=-\infty}^{\infty} e^{-\frac{|n|r\text{-}iny}{R}} = -\frac{G_{(4)}M}{4Rr}\left\{\frac{1}{1-e^{-\frac{r+iy}{R}}}+\frac{1}{1-e^{-\frac{r\text{-}iy}{R}}}-1\right\}, \quad (2.9)$$

where $r = |\mathbf{x}|$. Close to a source, where $z = |r + iy| \ll R$, the asymptotics is $\varphi(r,y) \to -\frac{1}{2}G_{(4)}M/z^2$ while at $r \gg R$ one has $\varphi(r,y) \to -GM/r$, i.e., the Newton potential with the effective gravitational constant

$$G \equiv G_{(3)} = \tfrac{1}{4}G_{(4)}/R. \quad (2.10)$$

## 3 The quantum Kepler problem in the compactified space

Let us consider the Schrödinger equation in the compactified space $\mathcal{M}_{3+1}$. Following the ADD hypothesis (Arkani-Hamed, *etel* 1998), the motion of a matter is constrained on the 3-brane $\mathfrak{R}^3 \subset \mathcal{M}_{3+1}$ which corresponds to the value $y = 0$ of the compact coordinate. The potential energy of the test particle of the mass $m$ on the brane is $U(r) = m\varphi(r,0)$, where $\varphi(r,y)$ is given in Eq. (2.9):

$$U(r) = -\frac{\alpha}{r}\text{cth}\left(\frac{r}{2R}\right) \quad (3.1)$$

where 
$$\alpha = GmM = \tfrac{1}{4}G_{(4)}mM/R \quad (3.2)$$

is a coupling constant. Asymptotics of the potential (3.1) at large and small distances follow from Subsection 2.3; they are

$$U_{(r \gg R)} = -\alpha/r, \qquad U_{(r \ll R)} = -2\alpha R/r^2. \quad (3.3)$$

A deviation from the Newton gravity law is notable at $r \leq R$. A singularity of the potential at $r \to 0$ leads to a drop of a particle in the center if its angular momentum $\mathbf{L}$ is relatively small but not necessarily zero:

$$\mathbf{L}^2 < 4\alpha mR \ (= 4m^2 MGR). \quad (3.4)$$

The drop in the center is an ill-posed peculiarity. It is unavoidable from the classical viewpoints but not from the quantum one. That is why we consider farther the quantum Kepler problem on the 3-brane $\Re^3 \subset \mathcal{M}_{3+1}$.

In order to analyze the Schrödinger equation with the potential (3.1) let us perform the radial reduction and introduce the dimensionless variables $\rho = r/a_g$, $\mathscr{E} = E/E_g$, where $a_g = \hbar^2/(m\alpha)$ and $E_g = m\alpha^2/\hbar^2 = \alpha/a_g$ are analogs of the Bohr radius and the Rydberg constant, and $E$ is the eigenenergy of the system. The equation for the radial wave function $\psi(\rho)$ is

$$\{H - \mathscr{E}\}\psi(\rho) = 0 \quad \text{with} \quad H = \frac{1}{2}\left\{-\frac{d^2}{d\rho^2} + \frac{\ell(\ell+1)}{\rho^2}\right\} + u(\rho), \tag{3.5}$$

$$u(\rho) = -\frac{1}{\rho}\operatorname{cth}\left(\frac{\rho}{2\delta}\right) = -\frac{1}{\rho}\left\{2\sum_{n=0}^{\infty} e^{-n\rho/\delta} - 1\right\}, \tag{3.6}$$

where $\ell = 0, 1, ...$ is the orbital quantum number and $\delta = R/a_g$. The exact solution of this equation is unknown, and we apply approximate methods.

## 3.1 Exact solutions in the lower-limiting Kratzer potential

Let us consider the sum of asymptotics (see Eq. (3.3)) of the potential (3.6):

$$\upsilon(\rho) \equiv u_{(\rho \gg \delta)} + u_{(\rho \ll \delta)} = -\frac{1}{\rho} - \frac{2\delta}{\rho^2}. \tag{3.7}$$

This is the Kratzer potential (Flügge 1971), but with the negative term $\sim 1/\rho^2$. The Kratzer problem is exactly solvable. It reduces to the Coulomb problem:

$$\{H_\lambda - \mathscr{E}\}\psi(\rho) = 0, \qquad H_\lambda = \frac{1}{2}\left\{-\frac{d^2}{d\rho^2} + \frac{\lambda(\lambda+1)}{\rho^2}\right\} - \frac{1}{\rho} \tag{3.8}$$

for the radially reduced Hamiltonian $H_\lambda$ and its eigenfunctions $\psi_{\lambda,n_r}(\rho)$ but with the noninteger orbital quantum number $\lambda = -\frac{1}{2} + \sqrt{(\ell+\frac{1}{2})^2 - 4\delta}$ (instead of $\ell = 0, 1, ...$). The spectrum of bound states is

$$\overline{\mathcal{E}}_{\ell,n_r} = -\frac{1}{2\nu^2}, \qquad \text{where} \quad \nu = n_r + \lambda + 1, \qquad (3.9)$$

and $n_r = 0,1,...$ (the radial quantum number). The potential (3.7) minorates the potential (3.6): $\upsilon(\rho) < u(\rho)$ for $\rho \in (0,\infty)$; Figure 3. Thus, one obtains the lower estimate for energy levels $\mathcal{E}_{\ell,n_r}$ of the Hamiltonian (3.5), (3.6): $\overline{\mathcal{E}}_{\ell,n_r} < \mathcal{E}_{\ell,n_r}$.

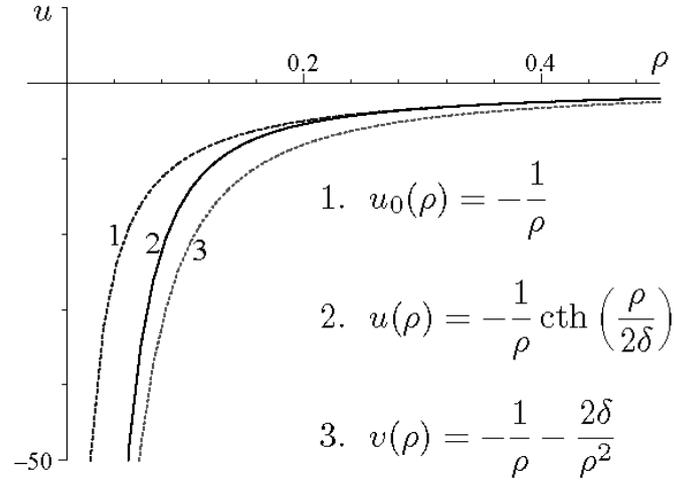

1. $u_0(\rho) = -\dfrac{1}{\rho}$
2. $u(\rho) = -\dfrac{1}{\rho}\operatorname{cth}\left(\dfrac{\rho}{2\delta}\right)$
3. $\upsilon(\rho) = -\dfrac{1}{\rho} - \dfrac{2\delta}{\rho^2}$

**Figure 3:** Behaviour of potentials for $\delta = 1/16$.

The potentials (3.6) and (3.7) have the same singularity $\sim -2\delta/\rho^2$ at $\rho \to 0$. Thus, both quantum problems are well posed provided

$$\delta \equiv m\alpha R/\hbar^2 = R/a_g < 1/16, \qquad (3.10)$$

otherwise some lower energy levels $\overline{\mathcal{E}}_{\ell,n_r}$ (as well as $\mathcal{E}_{\ell,n_r}$) become complex, and corresponding wave functions are not normalizable. Physically, this peculiarity corresponds to a drop in the center. In contrast to the classical problem, where the drop in the center is unavoidable for sufficiently small values of the angular momentum (3.4), quantum states are stable for arbitrary $\ell = 0,1,...$ provided the condition (3.10) holds.

## 3.2 The ground state via the variational method

The potential in Eq. (3.6) is expanded into a superposition of the Yukawa potentials. It is natural to apply in this case the variational approximation (Flügge, 1971) in order to derive an upper estimate for the ground state energy.

The ground state wave function is concentrated in the vicinity of the center where its properties are determined by the behaviour of the potential at $\rho \to 0$. In this area, the minoratig potential (3.7) simulates well the behaviour of the gravitational field – in contrast to the Newton potential; see Figure 3. Thus, it is appropriate, as a trial function, the scaled ground state wave function of the problem (3.8), i.e., $\widetilde{\psi}_{\lambda_0,0}(\rho) = \sqrt{\kappa}\psi_{\lambda_0,0}(\kappa\rho)$, where $\lambda_0 \equiv \lambda_{\ell=0} = \frac{1}{2}\left\{\sqrt{1-16\delta}-1\right\}$, and $\kappa$ is a variational parameter.

Integration techniques with hypergeometric functions (Landau & Lifshitz, 1981) lead to the expression for the average energy of the trial state:

$$\langle\mathscr{E}\rangle_\chi = \frac{\chi^2}{8(2v_0-1)\delta^2} - \frac{\chi}{2v_0\delta}\left\{2\chi^{2v_0}\zeta(2v_0,\chi)-1\right\}, \quad \text{where } v_0 = \lambda_0+1, \quad (3.11)$$

the new variational parameter $\chi = 2\kappa\delta/v_0$ is introduced for a conveniency (instead of $\kappa$), and $\zeta(\alpha,z)$ is the Hurwitz zeta function (Erdelyi, 1953):

$$\zeta(\alpha,z) = \sum_{n=0}^{\infty}(z+n)^{-\alpha}. \quad (3.12)$$

The minimum condition $\frac{\partial}{\partial\chi}\langle\mathscr{E}\rangle_\chi = 0$ for the energy (3.11) yields the equation

$$\chi = (1-v_0)(2v_0-1)\left\{(2v_0+1)\chi^{2v_0}\zeta(2v_0,\chi) - 2v_0\chi^{2v_0+1}\zeta(2v_0+1,\chi) - 1/2\right\}$$

which can be solved for $\chi$ numerically. In such a way, one obtains the dependency of $\chi$ on $v_0 = \frac{1}{2}\left\{\sqrt{1-16\delta}+1\right\}$ and thus on $\delta$. The substitution of the function $\chi(\delta)$ into Eq. (3.11) yields the dependency of the energy $\langle\mathscr{E}\rangle$ on $\delta$.

## 3.3 The ground state via the numerical integration

In order to estimate the precision of the variational approximation the problem (3.5), (3.6) was solved numerically, by means of the Runge-Kutta method, for different $\delta$. Results are presented in Figure 4 and Table 1. It is seen that, as $\delta$ grows, the ground state energy decreases slightly from $-½$ to $-0.7$ at $\delta = 1/16$. An error grows together with $\delta$ but does not exceed 3¼ %. Noteworthy, the lowest level $\overline{\mathcal{E}}_{0,0}$ of the problem (3.8) with the lower-limiting potential (3.7) is very crude estimate for the ground state energy (particularly, at $\delta = 1/16$), but the scaled eigenfunction $\widetilde{\psi}_{\lambda_0,0}(\rho)$ of this problem provides a satisfactory variational approximation for the problem (3.5), (3.6).

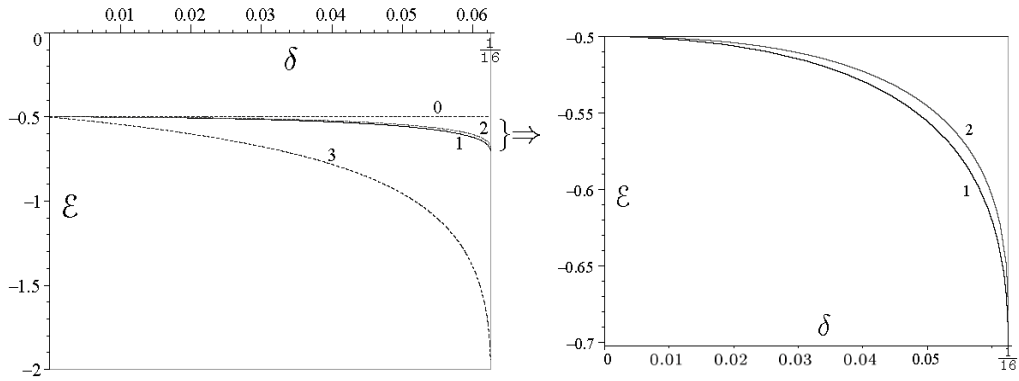

**Figure 4 & Table 1**: Ground state energy vs $\delta$, calculated via various methods: 1. numeral integration; 2. variational method; 3. lower-limiting potential; 4. perturbation method; 0. Coulomb ground state energy (for a comparison).

|   | $\delta$ | 0 | 1/64 | 1/32 | 3/64 | 1/16 |
|---|---|---|---|---|---|---|
| 1 | $\mathcal{E}_{num}$ | -½ | -0.50345517 | -0.51578916 | -0.54513697 | -0.70192749 |
| 2 | $\mathcal{E}_{var}$ | -½ | -0.50215711 | -0.51158195 | -0.53649797 | -0.67924964 |
| 3 | $\mathcal{E}_{min}$ | -½ | -0.57437416 | -0.68629150 | -8/9 | -2 |
| 4 | $\mathcal{E}_{per}$ | -½ | -0.50307198 | -0.51176891 | -0.52540078 | -0.54337917 |

(rows 1 and 2: $<3¼\%$)

## 3.4 Arbitrary states via the perturbation method

Wave functions of excited states extend far out the center as compared to one of the ground state. In this area, the potential (3.6) tends quickly to the Coulomb one; see Figure 3. In view of the condition $\delta < 1/16$, the parameter $\delta$ is small and thus it can serve as the expansion parameter in the perturbation method. Thus, the Kepler-Coulomb Hamiltonian with its eigenfunctions $\psi_{\ell, n_r}(\rho)$ should be used in zero-order approximation. The perturbation potential then is:

$$w(\rho) = -\frac{1}{\rho}\left\{\operatorname{cth}\left(\frac{\rho}{2\delta}\right) - 1\right\} = -\frac{2}{\rho}\sum_{k=1}^{\infty} e^{-k\rho/\delta}.$$

In the first-order approximation the energy of the state with quantum numbers $\ell, n_r$ reads $\mathscr{E}_{\ell, n_r} \approx \mathscr{E}_{\ell, n_r}^{(0)} + \mathscr{E}_{\ell, n_r}^{(1)}$, where

$$\mathscr{E}_{\ell, n_r}^{(0)} = -\frac{1}{2n^2} \quad \text{with} \quad n = n_r + \ell + 1 \quad \text{and} \quad \mathscr{E}_{\ell, n_r}^{(1)} = \langle w(\rho)\rangle_{\ell, n_r} \equiv \langle \ell, n_r | w(\rho) | \ell, n_r\rangle.$$

In order to evaluate the 1$^{\text{st}}$-order correction, one expresses $\psi_{\ell, n_r}(\rho)$ in terms of the confluent hypergeometric function and uses some techniques from (Landau & Lifshitz, 1981). The result is then represented in the form of finite sum:

$$\mathscr{E}_{\ell, n_r}^{(1)} = -\frac{2}{n^2}\sum_{j=0}^{2n_r}\sum_{i=0}^{[n_r - j/2]}(-)^j C_{n_r}^i C_{n+\ell}^{n_r - i} C_{2(n_r - i)}^j \left\{\varepsilon^{2n-j}\zeta(2n - j, \varepsilon) - 1\right\}, \qquad (3.13)$$

where the parameter $\varepsilon = 2\delta/n$ is as small as $\delta$, $C_l^k$ is the binomial coefficient, and [*expression*] in the upper limit of the sum stands for the integer part of *expression*.

Let us evaluate for the sum (3.13) few lower-order terms of the expansion in $\varepsilon$

$$\mathscr{E}_{\ell, n_r}^{(1)} = -\frac{2}{n^2} C_{n+\ell}^{n_r} \varepsilon^{2\ell+2}\left\{\zeta(2\ell + 2) - 2n\zeta(2\ell + 3)\varepsilon + \cdots\right\};$$

here $\zeta(z)$ is the Riemann zeta function (Erdelyi, 1953). As $\ell$ grows, the 1$^{\text{st}}$-order correction $\mathscr{E}_{\ell, n_r}^{(1)} = O(\varepsilon^{2\ell+2}) = O(\delta^{2\ell+2})$ decreases quickly, as expected.

But for the ground state the approximation is accurate for $\delta$ small, and even better than the variational approximation if $\delta \leq 0.03$; see Table 1. An energy correction in the 2$^\text{nd}$-order approximation of the perturbation method is not evaluated here but estimated as: $\mathcal{E}_{\ell,n_r}^{(2)} = O(\delta^{4\ell+4})$.

## 4 DISCUSSION

In the present paper, we have considerd two problems of the mathematical physics which are related to the hypothesis about existence in the space of extra compacts dimension.

The problem of a gravitation field generated by a pointlike mass in the compactified space of an arbitrary dimensionality is reduced to the multidimensional Poisson equation with the Dirac comb (or brush) source. A formal solution is expressed in terms of the Fourier integrals and series, the explicit form is derived for the spaces of 1+1 and 3+1 dimensions. The first example admits an electrotechnical treatment; the second one can be related to the nonrelativistic approximation of the Kaluza-Klein theory.

Short-range and long-range asymptotics of the potentials are consistent with those obtained earlier within the relativistic consideration. Moreover, the relation between the "genuine" gravitational constant in the multidimensional space and the effective gravitational constant observed far from the source (2.7) has the same nature as the relation between the Plank scale and electro-weak energy scale in ADD-hypothesis (Arkani-Hamed, *etel* 1998). Indeed, in terms of quantum units $c = \hbar = 1$ the gravitational constant in 3D space is $G \equiv G_{(3)} = 1/M_\text{Pl}^2$. Similarly, one can introduce for $G_{(3+d)}$ some mass $M_\text{f}$ such that $G_{(3+d)} = 1/M_\text{f}^{2+d}$. Up to numeral factors, the equality (2.7) yields for $D = 3$ the relation $M_{\text{Pl}}^2 \propto M_{\text{f}}^{2+d} R^d$ found in the Ref. (Arkani-Hamed, *etel* 1998) where $M_{\text{f}}$ is conjectured to be $M_{\text{f}} \approx M_{\text{ew}} \approx 1\,\text{TeV}$, i.e., the electroweak scale.

The second problem is the Kepler problem on 3D brane in 4D compactified space. It is shown that a classical motion in this case is unstable while the quantum version of the problem is self-consistent provided the constraint (3.10) holds. By now the physical meaning of the quantum Kepler problem is controversial since for all known particles it is impossible to identify gravitational effects in the background of other interactions. For example, for the gravitating electron + proton system the analogue of the Bohr radius $a_B$ is $a_g \approx 2 \cdot 10^{39}\, a_B$ which is bigger by many orders of the Universe extent, and a binding energy is negligibly small. The problem could be actual if there existed some dark matter superheavy particles of the mass $\geq 10^5$ TeV for which $a_g \leq 1$ mm (then it follows from (3.10) the condition $R \leq 0.1$ mm what is in accordance with modern estimates (Arkani-Hamed, *etel* 1998; Rubakov, 2001)). Another possibility is a hypothetical existence of non-gravitational sterile interactions penetrating into the extra dimensions (Rubakov, 2001).

A methodological interest to the quantum Kepler-Coulomb problem in the compactified space consists in the solution of the Schrödinger equation with new physically motivated potential. The influence of the extra compact dimension reduces to a lowering the energy of the ground state and weakly excited s-states by a quantity about $\Delta \mathscr{E} \propto \delta^2 = (R/a_g)^2$. More accurate results are derived by means of the perturbational, variational and numeral methods.

It is worth discussing the constraint (3.10) that is due to the singularity of the potential (3.1) at $r = 0$ within the nonrelativistic Schrödinger equation. One can argue that this constraint is meaningless since an actual relativistic motion is unstable because of another singularity on the Schwarzschild sphere. It turns out, however, that the constraint (3.10) has a relativistic implication. To show this, let us proceed from the following simple heuristic consideration.

It is known that the expression for the gravitational radius $r_g = 2GM/c^2$ can be derived within the Newtonian mechanics: $r_g$ is equal to the radius of the body of the mass $M$ for which the escape velocity $v$ is equal to the light speed $c$.

Similarly, one can obtain the gravitational radius $R_g$ in 4D space (infinite or compactified if $R_g < R$). Taking into account Eq. (2.3) for $N = 4$, the energy of a test particle of the mass $m$ in a gravity of the mass $M$ in this case is equal to:

$$E = \frac{mv^2}{2} - \frac{G_{(4)}mM}{2r^2}.$$

Conditions $E = 0$ and $v = c$ lead to the expression which coincides with the gravitational radius of a black hole in 4D space: $R_g = \sqrt{G_{(4)}M}/c$ (Myers & Perry, 1986).

It is known from the quantum field theory that any particle cannot be localized in a lesser extent than the Compton length $\lambdabar_C$. If $\lambdabar_C$ of a test particle is greater than the extent of a black hole,

$$\lambdabar_C \equiv \frac{\hbar}{mc} > 2R_g = 2\frac{\sqrt{G_{(4)}M}}{c}, \tag{4.1}$$

a vicinity (and the horizon in particular) of the black hole influences weakly a state of the particle. Thus, the constraint (4.1) is a natural necessary stability condition within the relativistic consideration. Noteworthy the light speed $c$ falls out the inequality (4.1) which upon Eqs. (2.10) and (3.2) reduces exactly to the constraint (3.10).